\documentclass[%
 reprint,
 amsmath,amssymb,
 aps,
]{revtex4-2}

\usepackage{graphicx}
\usepackage{dcolumn}
\usepackage{bm}

\begin{document}

\preprint{APS/123-QED}

\title{Mixing second- and third-order nonlinear interactions \\ in nanophotonic lithium-niobate waveguides}

\author{Simone Lauria}
 \email{sl196@hw.ac.uk}

 \affiliation{SUPA, Institute of Photonics and Quantum Sciences, Heriot-Watt University, EH14 4AS Edinburgh, UK}
\author{Mohammed F. Saleh}%
 \email{m.saleh@hw.ac.uk}
\affiliation{SUPA, Institute of Photonics and Quantum Sciences, Heriot-Watt University, EH14 4AS Edinburgh, UK}%




\date{\today}

\begin{abstract}
In this paper, we investigate  the interplay between the second and third-order nonlinearities in lithium-niobate waveguides with strong waveguide dispersion using uniform and linearly-chirped poling patterns at input powers in the pico-joule range. We implement the accurate unidirectional pulse propagation model to take into account all the possible nonlinear interactions inside these structures. In particular, the poling period is designed to quasi-phase-match single and multiple sum- and difference-frequency generation processes. We show how the poling period can be used as an additional degree of freedom to transform the output spectra of chip-based nonlinear waveguides in an unprecedented way.
\end{abstract}

\maketitle


\section{Introduction}
Nonlinear optical materials are classified into two main categories, centro- and non-centrosymmetric  \cite{Saleh07}. The former exhibits an inversion-symmetry that does not allow second-order nonlinear interactions to take place, whereas the dominance of the second-order nonlinearity in non-centrosymmetric materials usually results in overlooking third-order nonlinear effects.  With the rapid advances in fabrication techniques, strong third-order nonlinear interactions have been recently demonstrated in several structures, made of non-centrosymmetric materials, such as gallium-phosphide  bulk crystals and micro-resonators \cite{Rutkauskas20,Wilson20}, AlGaAs  waveguides and micro-resonators   \cite{Xie20,Kuyken20,Mahmudlu21}, and lithium-niobate, gallium-nitride and aluminium-nitride waveguides \cite{Stassen19,Lu20,Jankowski20}. Moreover, simultaneous second-order nonlinear interactions have also been demonstrated in few of these schemes, via satisfying the phase-matching condition using orientation-patterning \cite{Rutkauskas20} or periodic-poling techniques \cite{Jankowski20}. 

Lithium niobate (LiNbO$_3$) has always been considered the basis of the field of integrated optics because of its outstanding electro- and nonlinear-optical properties, its wide transparency window, as well as its compatibility with periodically-poled techniques   \cite{Jankowski20,Zhu21,Wang18}. Optical waveguides are usually obtained in LiNbO$_3$ substrates via  titanium indiffusion or proton-exchange. However, these structures offer weak confinement that limits their usage in nonlinear-optics applications with low-input powers. Recently, thin-film LiNbO$_3$ waveguides have been successfully bonded on top of silica substrates. This allows high contrast between the core and cladding. Subsequently, a drastic reduction of the operating pump level has been achieved.  

In this article, we will exploit  the poling period of the second-order nonlinear coefficient as an additional degree-of-freedom to allow for three-wave mixing between  different waves that comprise the supercontinuum  induced by the third-order nonlinear coefficient in LiNbO$_3$ waveguides with a silica substrate and pump power in the pico-joule range. We  show that the interplay between the concurrent second and third-order nonlinearities increases the functionality of optical devices, and  leads to the prediction  of new interesting phenomena.

The paper is organised as follows: the adopted model and the governing equations are introduced in Sec. II. Section III shows the simulations performed using uniform and linearly-chirped poled LiNbO$_3$ waveguides, and their various potential applications.  Finally, our conclusions are summarised in Sec. IV.


\section{Modelling second- and third-order nonlinear interactions in poled waveguides}
In this paper, we investigate the mixing between second- and third-order nonlinearities in the very-recently developed nanophotonic lithium-niobate waveguides with a strong mode confinement and very low-losses ($< 0.1$ dB/cm)\cite{Jankowski20}. In the latter reference, the adopted effective nonlinear Schr\"{o}dinger equation (NLSE), together with coupled-mode theory,  was unable to predict the experimentally observed higher-harmonic generation, as well as the effect of higher-order dispersion coefficients such as the dispersive-wave emission. Both effects play important roles in the presented study. Over the last decade, there have been very few numerical studies that have investigated the interplay between second- and third-order nonlinearities in  quadratic-poled germanium-doped photonic crystal fibres \cite{Baronio12} and lithium-niobate waveguides with weak confinement and relatively large field diameter \cite{Wabnitz10,Phillips11}. These studies were performed using the nonlinear envelope equation (NEE) \cite{Conforti10}. 

In this work, we implement  the unidirectional pulse propagation equation (UPPE) \cite{Kolesik04} that tracks the evolution of an optical pulse in nonlinear media more accurately. This model does not apply the slowly-varying approximation, a characteristic of the NLSE, hence, there are no limitations on the pulse spectral bandwidth. Unlike the NEE model, it includes the conjugated Kerr term \cite{Conforti13}, which results in the interaction with the negative frequency components of the pulse. We found that neglecting this term delays the spectral broadening by approximately 1 cm. Moreover, the UPPE model embeds the full nonlinear dispersion that is responsible for the self-steepening effect, essential in ultrashort-pulse dynamics. 

Consider the propagation of an ultrashort pulse in a waveguide with second and third-order nonlinearities. The spatial evolution of the pulse electric field using the UPPE model is governed by,
\begin{equation}
    \frac{\partial \Tilde{E}}{\partial z} = i [\beta (\omega) - \beta_1 \omega ] \Tilde{E} + i\frac{\omega ^2}{2 c^2 \epsilon _0 \beta (\omega)} \Tilde{P}_{\mathrm{NL}},\label{eq1}
\end{equation}
where $z$ is the propagation axis, $\omega$ is the angular frequency, $\tilde{E}(z,\omega) = \mathcal{F} \{E (z, t) \}$ is the spectral electric field, $t$ is the time in a reference frame moving with the pulse group velocity, $\mathcal{F}$ is the Fourier transform, $\beta (\omega)$ is the full dispersion, $\beta _1$ is the first-order dispersion coefficient, $c$ is the speed of light in vacuum, $\epsilon _0$ is the vacuum permittivity,  $ \tilde{P}_{\mathrm{NL}}(z, \omega) = \mathcal{F} \{ \epsilon_0 \chi^{(2)}(z) E^2(z, t) + \epsilon_0 \chi^{(3)} E^3(z, t) \}$ is the spectral nonlinear polarisation, and $\chi^{(2)}$ and $\chi^{(3)}$ are the second- and third-order nonlinear coefficients, respectively. It is worth to note that although $\chi^{(2)}$ can have a spatial dependence, $\chi^{(3)}$ is always uniform. The pulse envelope can be extracted using $A (z, t)=\mathcal{E} (z, t)e^{i [\omega_0 t-\beta (\omega_0)z]}$, with $\omega_0$  the pulse central frequency, $\mathcal{E}$ the analytical signal given by $\mathcal{E}(z, t)=E(z, t)-i\mathcal{H}\left[E(z, t)\right]$, and  $\mathcal{H}$ the Hilbert transform \cite{Conforti13}.

Periodic poling is an efficient technique to provide on-demand second-order nonlinear interactions. By flipping periodically the $\chi ^{(2)}$ coefficient along the direction of propagation at a certain rate, a specific parametric three-wave mixing can be significantly enhanced. To precisely and efficiently solve the UPPE in these media, we  implement an adaptive algorithm  based on the split-step Fourier method with a starting longitudinal step-size one-tenth of half the poling period \cite{Agrawal07}.  In this algorithm, if the energy is not conserved over a certain step, the step-size is halved and the propagation is recalculated. This process continues until the assigned condition is satisfied. 


Having a non-uniform poling-period pattern can enhance multiple interactions as the pump pulse propagates along the structure. This can result in an efficient up- or down-conversion process of a broadband spectrum. In this article, we study the nonlinear pulse propagation with uniform and linearly-chirped poling periods. For chirp structures, the poling period at any point along the propagation axis $z$ is given by,
\begin{equation}
    \frac{2 \pi}{ \Lambda (z)} = \frac{2\pi}{\Lambda _0} + \kappa z,
\end{equation}
where $\Lambda_0$ is the initial poling period at $z=0$, and $\kappa$ is the spatial frequency or the chirp parameter that can be positive or negative based on the initial and final poling periods. In our simulations, we approximate this continuous varying poling period via discretising the waveguide  into a relatively  large number of  segments with constant poling periods but different lengths in a staircase-like evolution, as depicted in Fig. 1(a). 

\begin{figure}[t]
    \includegraphics[width=\columnwidth]{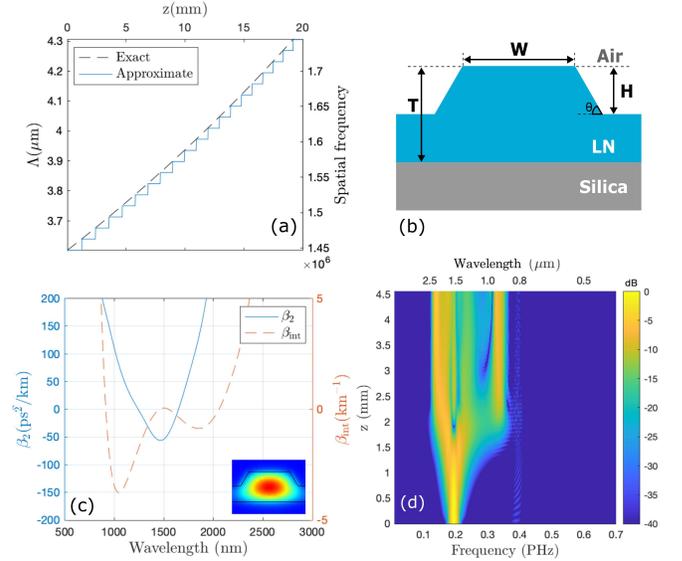}
    \caption{\label{fig:disp}(a) Spatial dependence of the poling period in the waveguide with an approximate linearly-chirped poling pattern. (b) A cross-section of a planar LiNbO$_3$ waveguide with 950 nm top width, 340 nm height, 800 nm thickness, 60$^{\mathrm{o}}$ angle, and silica bottom cladding. (c) The wavelength-dependence of the second-order dispersion $\beta_2$ (or group-velocity dispersion), and  the integrated dispersion $\beta_\mathrm{int}$, together with the fundamental TE-mode profile.  (d) Spectral evolution of the envelope of a pump source centred at 1550 nm, 5 pJ input energy, and 30 fs pulse duration through the waveguide without poling. As a guidance, the top axis is added to represent the corresponding wavelength. These pulse parameters are used in the rest of the simulations presented in this paper.}
    
\end{figure}

\section{Simulations and Discussions}
A cross-section of an x-cut LiNbO$_3$ waveguide that is used in the simulations presented in this work is displayed in Fig. \ref{fig:disp} (b). The waveguide has a trapezoidal core with 950 nm top width, 340 nm height, 800 nm thickness, a 60 degree angle, and a silica cladding substrate. Figure \ref{fig:disp} (c) shows the wavelength-dependence of the second-order dispersion coefficient $\beta_2$ of the fundamental  transverse-electric (TE) mode of the waveguide using a finite-element mode solver.  The material dispersion of LiNbO$_3$ and silica are  included via their Sellmeier equations \cite{Saleh07}. The waveguide operates in the anomalous dispersion regime in the telecommunication window. This would result in the generation of a broadband supercontinuum using a 1550 nm laser source. The integrated dispersion, defined as
\begin{equation}
\beta_{\mathrm{int}} (\omega) = \beta(\omega)-\beta (\omega_0)-\beta_1 (\omega_0)\left(\omega-\omega_0\right), 
\end{equation}
is also portrayed in Fig. \ref{fig:disp} (c). The zero crossings of $\beta_{\mathrm{int}}$ predict the wavelengths of the emitted dispersive waves during the supercontinuum generation \cite{Guo18}.

The spectral evolution of a 30 fs pulse, 5 pJ input energy, and 1550 nm central wavelength in a LiNbO$_3$ waveguide, with $\chi ^{(2)}=26$ pm/V, $\chi ^{(3)}=3417$ pm$^2$/V$^2$ (corresponding to nonlinear refractive index $2.6 \times 10^{-19} \mathrm{m}^2/\mathrm{W}$), and without poling, is displayed in Fig. \ref{fig:disp} (d).  As shown, the second-harmonic generation of the pump has very low efficiency due to the absence of a phase-matching technique. The supercontinuum generated due to Kerr nonlinearity spans from 900 nm to 2300 nm, and comprises the emission of two dispersive waves at 900 nm and 2050 nm, in very good agreement with the zero crossings of $\beta_{\mathrm{int}}$ displayed in Fig. \ref{fig:disp} (c). Linear losses are neglected here, since similar waveguides have reported very low losses \cite{Jankowski20}, particularly for the propagation lengths considered in this paper.


\subsection{Uniform poling}

\begin{figure}[t]
    \includegraphics[width=\columnwidth]{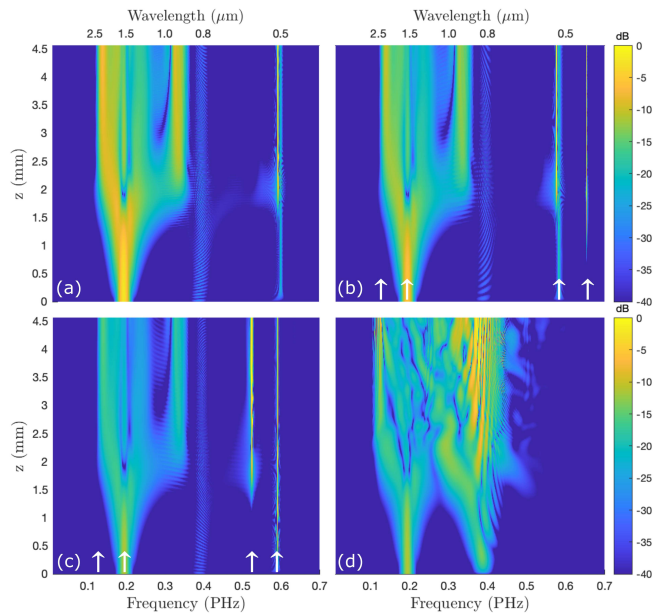}
    \caption{\label{fig:sfgs}Spectral evolution of the optical pulse envelope in the LiNbO$_3$ waveguide with four different uniform poling periods. (a) SFG between the pump and a wave at 738 nm, with $\Lambda=3.131\,\mu$m. (b) SFG between the pump and its second-harmonic with $\Lambda=3.261\,\mu$m, accompanied with a FWM process (highlighted by the white arrows) between the LDW, pump, THG, and an emitted wave at 450 nm. (c) SFG between the pump and SDW with $\Lambda=3.633\,\mu$m, accompanied with a FWM process (highlighted by the white arrows) between the LDW, pump, sum-frequency generated wave, and THG. (d) SFG between the pump and LDW with $\Lambda=4.285\,\mu$m. In all the subplots, the top axis is added to represent the corresponding wavelength, as a guidance.}
    
\end{figure}

We will first exploit the supercontinuum shown in Fig. \ref{fig:disp} (d) together with the uniform poling to showcase four different scenarios of the interplay between Type-0 sum-frequency generation (SFG) processes and the generated supercontinuum. To validate our numerical model, the poling period is selected to enhance the emission of a wave at 500 nm via a SFG process between the pump at 1550 nm and a wave at 738 nm, which results from the pulse broadening of the second-harmonic of the pump, using a poling period $\Lambda=3.131\, \mu$m. As displayed in Fig. \ref{fig:sfgs} (a), an up-converted wave is generated at 500 nm, then it subsequently merges with and enhances the third-harmonic at 517 nm. 

A third-harmonic wave can be weakly generated by coalescing  three pump waves in a third-order nonlinear medium, however, with very low efficiency due to the inherent phase mismatching. In Fig. \ref{fig:sfgs} (b), we exploit cascaded second-order nonlinear processes to directly enhance the third-harmonic generation (THG) that is seeded by third-order nonlinear processes. Using a poling period $\Lambda=3.261\,\mu$m, a SFG between the pump and its second harmonic is satisfied. Interestingly, another wave is also strongly emitted at 460 nm via a non-degenerate four-wave mixing (FWM) process with  the pump, the TH, and the longest-wavelength dispersive wave (LDW), such that its frequency is  $\omega_\mathrm{TH}+\omega_0-\omega_\mathrm{LDW}$. The FWM process is highlighted in Fig. \ref{fig:sfgs} (b)  by white arrows. 

The SFG between the pump and the shortest-wavelength dispersive wave (SDW) in a structure with $\Lambda=3.633 \,\mu$m results in  an up-converted wave at 575 nm, as  portrayed in Fig. \ref{fig:sfgs} (c).  Surprisingly, this configuration also indirectly enhances the THG via a FWM process with the LDW, the pump, and the up-converted wave with frequency $\omega_\mathrm{up}$, such that  $\omega_\mathrm{THG}=\omega_0+\omega_\mathrm{up}-\omega_\mathrm{LDW}$. It is important to note that this simultaneous FWM processes would not be predicted using the nonlinear Schr\"{o}dinger equation, similarly to the THG process.


The poling period in Fig. \ref{fig:sfgs} (d) is designed for an efficient  SFG between the pump and the LDW to amplify the SDW that lies close to the second harmonic. In this case, both the SDW and second harmonic are enhanced and  undergo self-broadening due to Kerr nonlinearity, then merge with the original spectrum to produce an ultra-broadband supercontinuum with very large intensity at the higher-frequency end of the spectrum. This configuration shows the potential of these new integrated devices for novel optical applications in the visible and the near-visible regions using the waveguide fundamental mode. In these regimes, efficient third-order nonlinear interactions become very challenging,  because the group-velocity dispersion is usually highly normal at the fundamental mode and therefore higher-order mode operation is sought instead \cite{Zhao20}.

\begin{figure}[t]
    \includegraphics[width=\columnwidth]{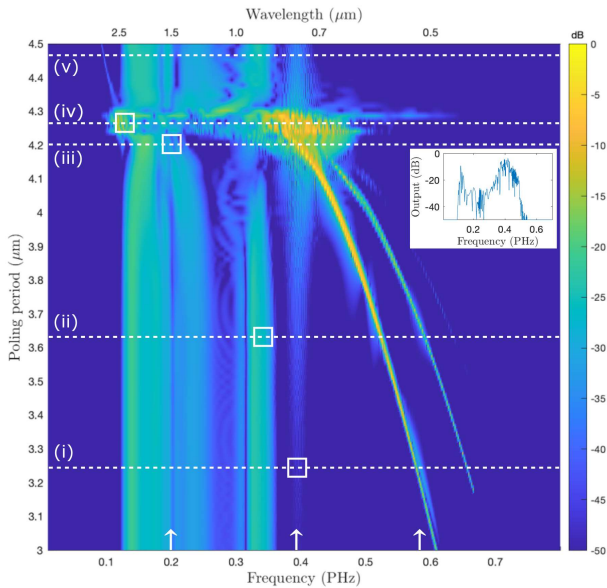}
    \caption{\label{fig:summary} Output spectra of the optical pulse envelope in the LiNbO$_3$ waveguide and its dependence on the poling period. The color plot is normalised to its maximum. White dashed lines correspond to: (i) THG via two cascaded SFG processes [Fig. \ref{fig:sfgs} (b)]; (ii) THG via a SFG followed by a FWM process [Fig. \ref{fig:sfgs} (c)]; (iii) Second-harmonic generation; (iv) Ultra-broadband supercontinuum; (v) Difference-frequency generation. White squares represent the interacting waves with the pump via the second-order nonlinearity of the medium. The inset depicts the output spectrum normalised to its maximum at $\Lambda = 4.25\, \mu$m.}
    
\end{figure}

Additional insights on the dependence of the features of the output spectra of the waveguide on the poling period can be elicited from Fig. \ref{fig:summary}.  The white dashed lines represent different scenarios.  The interacting wave with the pump through the second-order nonlinearity is indicated by  white squares. The frequency of this interacting wave is down shifted from the blue side of the spectrum as the poling period increases. Subsequently, the wavelengths of the up-converted waves induced by the SFG and the accompanied FWM processes are shifted downwards. Relatively shorter periods can be exploited to enhance the THG via either two cascaded SFG processes [case (i)], or a SFG interaction followed by a FWM process [case (ii)]. The second-harmonic is strongly enhanced with $\Lambda = 4.2\, \mu$m that allows efficient doubling of the frequency of the pump accompanied with strong depletion [case (iii)]. Slightly increasing the poling period beyond the latter value, we approach an interesting regime where an ultra-broadband spectrum is obtained via allowing a SFG between the pump and the LDW to booster the SDW emission [case (iv)]. In this case the output power of the blue side of the spectrum is very intense, in comparison to the rest of the plot. The shown inset validates the numerical simulations at these particular range of poling periods. Increasing the poling period beyond that limit, a difference-frequency generation process with mid-infra red emission starts to take place, as discussed below in more details [case (v)].

\subsection{Linearly-chirped poling}

\begin{figure}
    \includegraphics[width=\columnwidth]{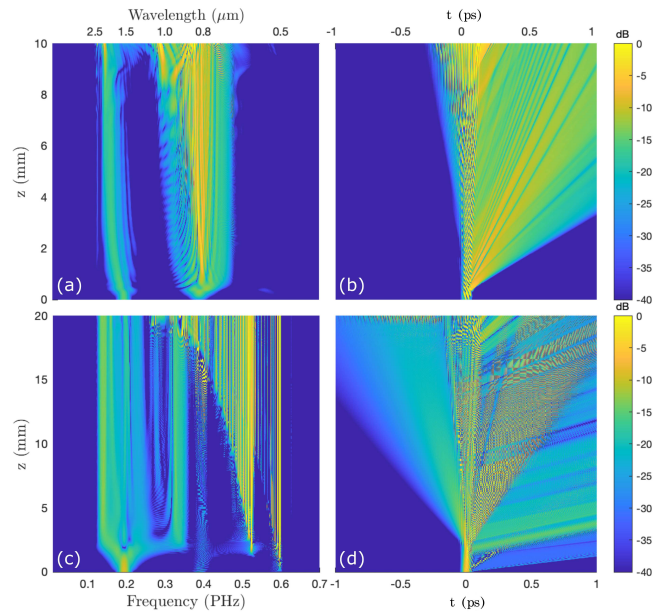}
    \caption{\label{fig:chirp}Spectral and temporal evolutions of the optical pulse envelope in the LiNbO$_3$ waveguide with two different linearly-chirped poling periods for SFG processes. (a,b) $\Lambda_0 = 4.199\,\mu$m and $\Lambda_\mathrm{L} = 4.306\,\mu$m. Poling period varies in 10 steps. (c,d) $\Lambda_0 = 3.602\,\mu$m and $\Lambda_\mathrm{L} = 4.306\,\mu$m. Poling period varies in 20 steps.}
    
\end{figure}

Using a chirp-poling pattern can satisfy the phase-matching conditions of multiple second-order nonlinear interactions. The propagation of the aforementioned optical pulse in linearly-chirped poled LiNbO$_3$ waveguides is depicted in Fig. \ref{fig:chirp} with two different chirp parameters. The poling period used in panels (a,b) is gradually increased in 10 segments from $\Lambda_0 = 4.199\, \mu$m to $\Lambda_\mathrm{L} = \Lambda(z=L) = 4.306\, \mu $m through a 10 mm long waveguide. These poling periods allow for a sequence of efficient SFG processes between the pump and the part of the spectrum that spans from 1500 nm to 2300 nm. As shown, a large portion of the original spectrum is successfully up-converted, broadens, and significantly enhances the supercontinuum generation around the  second-harmonic in the visible region. This is accompanied with a pulse temporal broadening as shown in Fig. \ref{fig:chirp} (b).

The spectral and temporal evolutions of the pulse in a 20 mm-long LiNbO$_3$ waveguide, where the  poling period increases in 20 segments from $3.602\, \mu $m to $4.306 \,\mu $m are displayed in Figs. \ref{fig:chirp} (c,d). This range of poling periods enhances  SFG between the pump and the entire supercontinuum, from 900 nm to 2300 nm over a reasonable interaction length. Interestingly, this scheme is able to generate a train of narrowband pulses that are nearly equally spaced in the visible regime, approaching a comb-light generation. The origin of this set is different from the frequency comb inherent of supercontinuum generation, with frequency-spacing determined by the repetition rate of the pump source. This set of pulses is due to the approximately equally-spaced spatial frequencies provided by the linearly-chirped poling pattern, shown in Fig. \ref{fig:disp} (a), used in correcting phase mismatching between the interacting waves. Moreover, another set of nearly equally spaced waves are generated at higher frequencies via multiple FWM processes with the pump, the LDW, and the former set of waves. Similar to the dynamics using uniform poling, we find by increasing the initial poling period or decreasing the chirp parameter these two sets of waves are shifted downwards following the interacting waves with the pump. 

\begin{figure}[t]
    \includegraphics[width=\columnwidth]{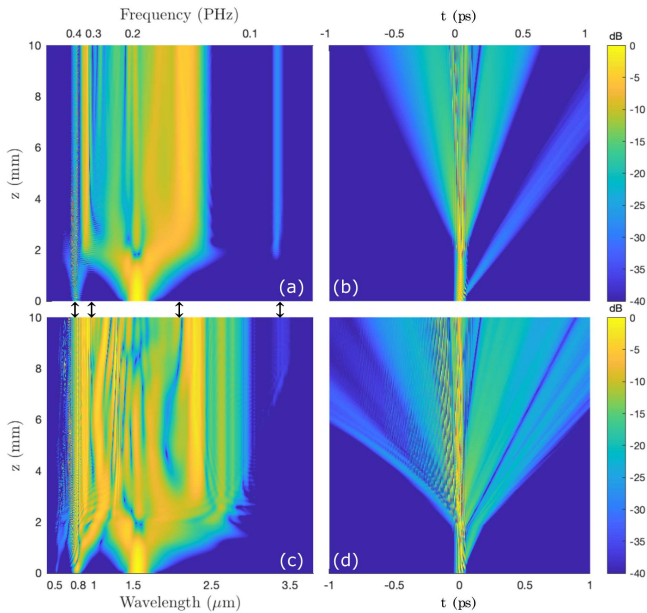}
    \caption{\label{fig:dfg}Spectral and temporal evolutions of the optical pulse envelope in the LiNbO$_3$ waveguide with two different linearly-chirped poling periods for DFG processes. (a,b) $\Lambda_0 = 4.506\,\mu$m and $\Lambda_\mathrm{L} = 4.723\,\mu$m. Poling period varies in 10 steps. (c,d) $\Lambda_0 = 4.278\,\mu$m and  $\Lambda_\mathrm{L} = 4.723\,\mu$m. Poling period varies in 15 steps.}    
\end{figure}

The poling period could also be exploited to down-convert the original spectrum induced by Kerr nonlinearity to the mid-infrared (MIR) regime by allowing multiple difference-frequency generation (DFG) processes to take place efficiently. The spectral and temporal evolutions of the pulse in a LiNbO$_3$ waveguide, where the  poling period increases in 10 steps from $4.506 \, \mu $m to $4.723 \,\mu $m, are shown in Figs. \ref{fig:dfg} (a,b). This waveguide is supposed to down-convert a small portion of the spectrum, in the range 1050--1100 nm, to around 3200--3800 nm via DFG processes with the pump. However we find that because of the competition with the SFG processes, only a narrowband wave at $3.35 \, \mu $m is emitted. The SFG processes are more favourable than the DFG processes, because the former are usually seeded by high harmonics. Interestingly, this wave at $3.35 \, \mu $m is also concurrently seeded via a FWM process with the LDW, SDW, and the second harmonic (SH), such that its frequency is governed by $\omega_\mathrm{LDW}+\omega_\mathrm{SDW}-\omega_\mathrm{SH}$. This regime of operation is represented by the upper part of Fig. \ref{fig:summary}, with $\Lambda_\mathrm{L} > 4.3\,\mu$m.

Panels (c,d) show the evolution of the pump in the  waveguide with 15 different poling periods, between $4.278 \, \mu $m and $4.723 \, \mu $m, that support multiple DFG processes between the pump and waves in the range 875--1100 nm. This scheme should result in the emission of  waves in the MIR regime  from $\approx 2 \,\mu$m to $3.8\,\mu$m. However, the DFG processes only strengthen and extend the red-side of the spectrum to approximately $3.2\,\mu$m, as well as the FWM process at $3.35\,\mu$m. Similarly to the previous case, no waves are observed beyond that limit because of the competition with SFG processes. The shortest poling period of this structure is also close to the one used for the second-harmonic generation ($4.214 \, \mu $m), hence, the latter gets amplified, broadened, coalesced with the original spectrum, and  generates an ultra-broadband supercontinuum that extends from the visible to the MIR regime.

\section{Conclusions}
In conclusion, we show the potential of the interplay between the second- and third-order nonlinearities in optical  waveguides with uniform and linearly-chirped poling patterns in transforming the output spectra. This interplay is studied using the accurate UPPE model that takes into account all the possible nonlinear interactions.  The poling patterns are designed to quasi-phase-match single or multiple sets of three waves within the pulse spectrum, originally induced  by Kerr nonlinearity. We find that this three-wave mixing process is usually accompanied by another FWM process that also involves the strongly-emitted dispersive waves.  This results in various outputs, such as ultra-broad supercontinuum, two coupled sets of narrowband pulses that are nearly equally spaced, and efficient up-conversion to the  visible regime, all at a fundamental-mode operation. The latter application can be considered as an alternative potential technique for higher-mode operation to achieve efficient third-order nonlinear interactions in the visible region. The simulated structures can be feasibly fabricated using the current technologies, which would facilitate the experimental demonstration of this work in the very near future. 

Finally, this research shows the endless possibilities  of altering the output spectra in novel waveguides using designed poling periods of the second-order nonlinear coefficient. This will undoubtedly provide fruitful opportunities to advance the field of chip-based nonlinear photonics.

\section*{Acknowledgement} This research is supported by EPSRC Doctoral Training Partnerships (DTP) programme EP/R513040/1.




\begin{thebibliography}{21}%
\makeatletter
\providecommand \@ifxundefined [1]{%
 \@ifx{#1\undefined}
}%
\providecommand \@ifnum [1]{%
 \ifnum #1\expandafter \@firstoftwo
 \else \expandafter \@secondoftwo
 \fi
}%
\providecommand \@ifx [1]{%
 \ifx #1\expandafter \@firstoftwo
 \else \expandafter \@secondoftwo
 \fi
}%
\providecommand \natexlab [1]{#1}%
\providecommand \enquote  [1]{``#1''}%
\providecommand \bibnamefont  [1]{#1}%
\providecommand \bibfnamefont [1]{#1}%
\providecommand \citenamefont [1]{#1}%
\providecommand \href@noop [0]{\@secondoftwo}%
\providecommand \href [0]{\begingroup \@sanitize@url \@href}%
\providecommand \@href[1]{\@@startlink{#1}\@@href}%
\providecommand \@@href[1]{\endgroup#1\@@endlink}%
\providecommand \@sanitize@url [0]{\catcode `\\12\catcode `\$12\catcode
  `\&12\catcode `\#12\catcode `\^12\catcode `\_12\catcode `\%12\relax}%
\providecommand \@@startlink[1]{}%
\providecommand \@@endlink[0]{}%
\providecommand \url  [0]{\begingroup\@sanitize@url \@url }%
\providecommand \@url [1]{\endgroup\@href {#1}{\urlprefix }}%
\providecommand \urlprefix  [0]{URL }%
\providecommand \Eprint [0]{\href }%
\providecommand \doibase [0]{https://doi.org/}%
\providecommand \selectlanguage [0]{\@gobble}%
\providecommand \bibinfo  [0]{\@secondoftwo}%
\providecommand \bibfield  [0]{\@secondoftwo}%
\providecommand \translation [1]{[#1]}%
\providecommand \BibitemOpen [0]{}%
\providecommand \bibitemStop [0]{}%
\providecommand \bibitemNoStop [0]{.\EOS\space}%
\providecommand \EOS [0]{\spacefactor3000\relax}%
\providecommand \BibitemShut  [1]{\csname bibitem#1\endcsname}%
\let\auto@bib@innerbib\@empty
\bibitem [{\citenamefont {Saleh}\ and\ \citenamefont {Teich}(2007)}]{Saleh07}%
  \BibitemOpen
  \bibfield  {author} {\bibinfo {author} {\bibfnamefont {B.~E.~A.}\
  \bibnamefont {Saleh}}\ and\ \bibinfo {author} {\bibfnamefont {M.~C.}\
  \bibnamefont {Teich}},\ }\href@noop {} {\emph {\bibinfo {title} {Fundamentals
  of Photonics}}},\ \bibinfo {edition} {2nd}\ ed.\ (\bibinfo  {publisher} {John
  Wiley \& Sons, Hoboken, New Jersey},\ \bibinfo {year} {2007})\BibitemShut
  {NoStop}%
\bibitem [{\citenamefont {Rutkauskas}\ \emph {et~al.}(2020)\citenamefont
  {Rutkauskas}, \citenamefont {Srivastava},\ and\ \citenamefont
  {Reid}}]{Rutkauskas20}%
  \BibitemOpen
  \bibfield  {author} {\bibinfo {author} {\bibfnamefont {M.}~\bibnamefont
  {Rutkauskas}}, \bibinfo {author} {\bibfnamefont {A.}~\bibnamefont
  {Srivastava}},\ and\ \bibinfo {author} {\bibfnamefont {D.~T.}\ \bibnamefont
  {Reid}},\ }\bibfield  {title} {\bibinfo {title} {Supercontinuum generation in
  orientation-patterned gallium phosphide},\ }\href
  {https://doi.org/10.1364/OPTICA.385200} {\bibfield  {journal} {\bibinfo
  {journal} {Optica}\ }\textbf {\bibinfo {volume} {7}},\ \bibinfo {pages} {172}
  (\bibinfo {year} {2020})}\BibitemShut {NoStop}%
\bibitem [{\citenamefont {Wilson}\ \emph {et~al.}(2020)\citenamefont {Wilson},
  \citenamefont {Schneider}, \citenamefont {H{\"o}nl}, \citenamefont
  {Anderson}, \citenamefont {Baumgartner}, \citenamefont {Czornomaz},
  \citenamefont {Kippenberg},\ and\ \citenamefont {Seidler}}]{Wilson20}%
  \BibitemOpen
  \bibfield  {author} {\bibinfo {author} {\bibfnamefont {D.~J.}\ \bibnamefont
  {Wilson}}, \bibinfo {author} {\bibfnamefont {K.}~\bibnamefont {Schneider}},
  \bibinfo {author} {\bibfnamefont {S.}~\bibnamefont {H{\"o}nl}}, \bibinfo
  {author} {\bibfnamefont {M.}~\bibnamefont {Anderson}}, \bibinfo {author}
  {\bibfnamefont {Y.}~\bibnamefont {Baumgartner}}, \bibinfo {author}
  {\bibfnamefont {L.}~\bibnamefont {Czornomaz}}, \bibinfo {author}
  {\bibfnamefont {T.~J.}\ \bibnamefont {Kippenberg}},\ and\ \bibinfo {author}
  {\bibfnamefont {P.}~\bibnamefont {Seidler}},\ }\bibfield  {title} {\bibinfo
  {title} {Integrated gallium phosphide nonlinear photonics},\ }\href
  {https://doi.org/10.1038/s41566-019-0537-9} {\bibfield  {journal} {\bibinfo
  {journal} {Nature Photonics}\ }\textbf {\bibinfo {volume} {14}},\ \bibinfo
  {pages} {57} (\bibinfo {year} {2020})}\BibitemShut {NoStop}%
\bibitem [{\citenamefont {Xie}\ \emph {et~al.}(2020)\citenamefont {Xie},
  \citenamefont {Chang}, \citenamefont {Shu}, \citenamefont {Norman},
  \citenamefont {Peters}, \citenamefont {Wang},\ and\ \citenamefont
  {Bowers}}]{Xie20}%
  \BibitemOpen
  \bibfield  {author} {\bibinfo {author} {\bibfnamefont {W.}~\bibnamefont
  {Xie}}, \bibinfo {author} {\bibfnamefont {L.}~\bibnamefont {Chang}}, \bibinfo
  {author} {\bibfnamefont {H.}~\bibnamefont {Shu}}, \bibinfo {author}
  {\bibfnamefont {J.~C.}\ \bibnamefont {Norman}}, \bibinfo {author}
  {\bibfnamefont {J.~D.}\ \bibnamefont {Peters}}, \bibinfo {author}
  {\bibfnamefont {X.}~\bibnamefont {Wang}},\ and\ \bibinfo {author}
  {\bibfnamefont {J.~E.}\ \bibnamefont {Bowers}},\ }\bibfield  {title}
  {\bibinfo {title} {Ultrahigh-q algaas-on-insulator microresonators for
  integrated nonlinear photonics},\ }\href {https://doi.org/10.1364/OE.405343}
  {\bibfield  {journal} {\bibinfo  {journal} {Opt. Express}\ }\textbf {\bibinfo
  {volume} {28}},\ \bibinfo {pages} {32894} (\bibinfo {year}
  {2020})}\BibitemShut {NoStop}%
\bibitem [{\citenamefont {Kuyken}\ \emph {et~al.}(2020)\citenamefont {Kuyken},
  \citenamefont {Billet}, \citenamefont {Leo}, \citenamefont {Yvind},\ and\
  \citenamefont {Pu}}]{Kuyken20}%
  \BibitemOpen
  \bibfield  {author} {\bibinfo {author} {\bibfnamefont {B.}~\bibnamefont
  {Kuyken}}, \bibinfo {author} {\bibfnamefont {M.}~\bibnamefont {Billet}},
  \bibinfo {author} {\bibfnamefont {F.}~\bibnamefont {Leo}}, \bibinfo {author}
  {\bibfnamefont {K.}~\bibnamefont {Yvind}},\ and\ \bibinfo {author}
  {\bibfnamefont {M.}~\bibnamefont {Pu}},\ }\bibfield  {title} {\bibinfo
  {title} {Octave-spanning coherent supercontinuum generation in an
  algaas-on-insulator waveguide},\ }\href
  {https://doi.org/10.1364/OL.45.000603} {\bibfield  {journal} {\bibinfo
  {journal} {Opt. Lett.}\ }\textbf {\bibinfo {volume} {45}},\ \bibinfo {pages}
  {603} (\bibinfo {year} {2020})}\BibitemShut {NoStop}%
\bibitem [{\citenamefont {Mahmudlu}\ \emph {et~al.}(2021)\citenamefont
  {Mahmudlu}, \citenamefont {May}, \citenamefont {Angulo}, \citenamefont
  {Sorel},\ and\ \citenamefont {Kues}}]{Mahmudlu21}%
  \BibitemOpen
  \bibfield  {author} {\bibinfo {author} {\bibfnamefont {H.}~\bibnamefont
  {Mahmudlu}}, \bibinfo {author} {\bibfnamefont {S.}~\bibnamefont {May}},
  \bibinfo {author} {\bibfnamefont {A.}~\bibnamefont {Angulo}}, \bibinfo
  {author} {\bibfnamefont {M.}~\bibnamefont {Sorel}},\ and\ \bibinfo {author}
  {\bibfnamefont {M.}~\bibnamefont {Kues}},\ }\bibfield  {title} {\bibinfo
  {title} {Algaas-on-insulator waveguide for highly efficient photon-pair
  generation via spontaneous four-wave mixing},\ }\href
  {https://doi.org/10.1364/OL.418932} {\bibfield  {journal} {\bibinfo
  {journal} {Opt. Lett.}\ }\textbf {\bibinfo {volume} {46}},\ \bibinfo {pages}
  {1061} (\bibinfo {year} {2021})}\BibitemShut {NoStop}%
\bibitem [{\citenamefont {Stassen}\ \emph {et~al.}(2019)\citenamefont
  {Stassen}, \citenamefont {Pu}, \citenamefont {Semenova}, \citenamefont
  {Zavarin}, \citenamefont {Lundin},\ and\ \citenamefont {Yvind}}]{Stassen19}%
  \BibitemOpen
  \bibfield  {author} {\bibinfo {author} {\bibfnamefont {E.}~\bibnamefont
  {Stassen}}, \bibinfo {author} {\bibfnamefont {M.}~\bibnamefont {Pu}},
  \bibinfo {author} {\bibfnamefont {E.}~\bibnamefont {Semenova}}, \bibinfo
  {author} {\bibfnamefont {E.}~\bibnamefont {Zavarin}}, \bibinfo {author}
  {\bibfnamefont {W.}~\bibnamefont {Lundin}},\ and\ \bibinfo {author}
  {\bibfnamefont {K.}~\bibnamefont {Yvind}},\ }\bibfield  {title} {\bibinfo
  {title} {High-confinement gallium nitride-on-sapphire waveguides for
  integrated nonlinear photonics},\ }\href
  {https://doi.org/10.1364/OL.44.001064} {\bibfield  {journal} {\bibinfo
  {journal} {Opt. Lett.}\ }\textbf {\bibinfo {volume} {44}},\ \bibinfo {pages}
  {1064} (\bibinfo {year} {2019})}\BibitemShut {NoStop}%
\bibitem [{\citenamefont {Lu}\ \emph {et~al.}(2020)\citenamefont {Lu},
  \citenamefont {Liu}, \citenamefont {Bruch}, \citenamefont {Zhang},
  \citenamefont {Wang}, \citenamefont {Yan},\ and\ \citenamefont
  {Tang}}]{Lu20}%
  \BibitemOpen
  \bibfield  {author} {\bibinfo {author} {\bibfnamefont {J.}~\bibnamefont
  {Lu}}, \bibinfo {author} {\bibfnamefont {X.}~\bibnamefont {Liu}}, \bibinfo
  {author} {\bibfnamefont {A.~W.}\ \bibnamefont {Bruch}}, \bibinfo {author}
  {\bibfnamefont {L.}~\bibnamefont {Zhang}}, \bibinfo {author} {\bibfnamefont
  {J.}~\bibnamefont {Wang}}, \bibinfo {author} {\bibfnamefont {J.}~\bibnamefont
  {Yan}},\ and\ \bibinfo {author} {\bibfnamefont {H.~X.}\ \bibnamefont
  {Tang}},\ }\bibfield  {title} {\bibinfo {title} {Ultraviolet to mid-infrared
  supercontinuum generation in single-crystalline aluminum nitride
  waveguides},\ }\href {https://doi.org/10.1364/OL.398257} {\bibfield
  {journal} {\bibinfo  {journal} {Opt. Lett.}\ }\textbf {\bibinfo {volume}
  {45}},\ \bibinfo {pages} {4499} (\bibinfo {year} {2020})}\BibitemShut
  {NoStop}%
\bibitem [{\citenamefont {Jankowski}\ \emph {et~al.}(2020)\citenamefont
  {Jankowski}, \citenamefont {Langrock}, \citenamefont {Desiatov},
  \citenamefont {Marandi}, \citenamefont {Wang}, \citenamefont {Zhang},
  \citenamefont {Phillips}, \citenamefont {Lon\v{c}ar},\ and\ \citenamefont
  {Fejer}}]{Jankowski20}%
  \BibitemOpen
  \bibfield  {author} {\bibinfo {author} {\bibfnamefont {M.}~\bibnamefont
  {Jankowski}}, \bibinfo {author} {\bibfnamefont {C.}~\bibnamefont {Langrock}},
  \bibinfo {author} {\bibfnamefont {B.}~\bibnamefont {Desiatov}}, \bibinfo
  {author} {\bibfnamefont {A.}~\bibnamefont {Marandi}}, \bibinfo {author}
  {\bibfnamefont {C.}~\bibnamefont {Wang}}, \bibinfo {author} {\bibfnamefont
  {M.}~\bibnamefont {Zhang}}, \bibinfo {author} {\bibfnamefont {C.~R.}\
  \bibnamefont {Phillips}}, \bibinfo {author} {\bibfnamefont {M.}~\bibnamefont
  {Lon\v{c}ar}},\ and\ \bibinfo {author} {\bibfnamefont {M.~M.}\ \bibnamefont
  {Fejer}},\ }\bibfield  {title} {\bibinfo {title} {Ultrabroadband nonlinear
  optics in nanophotonic periodically poled lithium niobate waveguides},\
  }\href {https://doi.org/10.1364/OPTICA.7.000040} {\bibfield  {journal}
  {\bibinfo  {journal} {Optica}\ }\textbf {\bibinfo {volume} {7}},\ \bibinfo
  {pages} {40} (\bibinfo {year} {2020})}\BibitemShut {NoStop}%
\bibitem [{\citenamefont {Zhu}\ \emph {et~al.}(2021)\citenamefont {Zhu},
  \citenamefont {Shao}, \citenamefont {Yu}, \citenamefont {Cheng},
  \citenamefont {Desiatov}, \citenamefont {Xin}, \citenamefont {Hu},
  \citenamefont {Holzgrafe}, \citenamefont {Ghosh}, \citenamefont
  {Shams-Ansari}, \citenamefont {Puma}, \citenamefont {Sinclair}, \citenamefont
  {Reimer}, \citenamefont {Zhang},\ and\ \citenamefont {Lon\v{c}ar}}]{Zhu21}%
  \BibitemOpen
  \bibfield  {author} {\bibinfo {author} {\bibfnamefont {D.}~\bibnamefont
  {Zhu}}, \bibinfo {author} {\bibfnamefont {L.}~\bibnamefont {Shao}}, \bibinfo
  {author} {\bibfnamefont {M.}~\bibnamefont {Yu}}, \bibinfo {author}
  {\bibfnamefont {R.}~\bibnamefont {Cheng}}, \bibinfo {author} {\bibfnamefont
  {B.}~\bibnamefont {Desiatov}}, \bibinfo {author} {\bibfnamefont {C.~J.}\
  \bibnamefont {Xin}}, \bibinfo {author} {\bibfnamefont {Y.}~\bibnamefont
  {Hu}}, \bibinfo {author} {\bibfnamefont {J.}~\bibnamefont {Holzgrafe}},
  \bibinfo {author} {\bibfnamefont {S.}~\bibnamefont {Ghosh}}, \bibinfo
  {author} {\bibfnamefont {A.}~\bibnamefont {Shams-Ansari}}, \bibinfo {author}
  {\bibfnamefont {E.}~\bibnamefont {Puma}}, \bibinfo {author} {\bibfnamefont
  {N.}~\bibnamefont {Sinclair}}, \bibinfo {author} {\bibfnamefont
  {C.}~\bibnamefont {Reimer}}, \bibinfo {author} {\bibfnamefont
  {M.}~\bibnamefont {Zhang}},\ and\ \bibinfo {author} {\bibfnamefont
  {M.}~\bibnamefont {Lon\v{c}ar}},\ }\bibfield  {title} {\bibinfo {title}
  {Integrated photonics on thin-film lithium niobate},\ }\href
  {https://doi.org/10.1364/AOP.411024} {\bibfield  {journal} {\bibinfo
  {journal} {Adv. Opt. Photon.}\ }\textbf {\bibinfo {volume} {13}},\ \bibinfo
  {pages} {242} (\bibinfo {year} {2021})}\BibitemShut {NoStop}%
\bibitem [{\citenamefont {Wang}\ \emph {et~al.}(2018)\citenamefont {Wang},
  \citenamefont {Langrock}, \citenamefont {Marandi}, \citenamefont {Jankowski},
  \citenamefont {Zhang}, \citenamefont {Desiatov}, \citenamefont {Fejer},\ and\
  \citenamefont {Lon\v{c}ar}}]{Wang18}%
  \BibitemOpen
  \bibfield  {author} {\bibinfo {author} {\bibfnamefont {C.}~\bibnamefont
  {Wang}}, \bibinfo {author} {\bibfnamefont {C.}~\bibnamefont {Langrock}},
  \bibinfo {author} {\bibfnamefont {A.}~\bibnamefont {Marandi}}, \bibinfo
  {author} {\bibfnamefont {M.}~\bibnamefont {Jankowski}}, \bibinfo {author}
  {\bibfnamefont {M.}~\bibnamefont {Zhang}}, \bibinfo {author} {\bibfnamefont
  {B.}~\bibnamefont {Desiatov}}, \bibinfo {author} {\bibfnamefont {M.~M.}\
  \bibnamefont {Fejer}},\ and\ \bibinfo {author} {\bibfnamefont
  {M.}~\bibnamefont {Lon\v{c}ar}},\ }\bibfield  {title} {\bibinfo {title}
  {Ultrahigh-efficiency wavelength conversion in nanophotonic periodically
  poled lithium niobate waveguides},\ }\href
  {https://doi.org/10.1364/OPTICA.5.001438} {\bibfield  {journal} {\bibinfo
  {journal} {Optica}\ }\textbf {\bibinfo {volume} {5}},\ \bibinfo {pages}
  {1438} (\bibinfo {year} {2018})}\BibitemShut {NoStop}%
\bibitem [{\citenamefont {Baronio}\ \emph {et~al.}(2012)\citenamefont
  {Baronio}, \citenamefont {Conforti}, \citenamefont {{De Angelis}},
  \citenamefont {Modotto}, \citenamefont {Wabnitz}, \citenamefont {Andreana},
  \citenamefont {Tonello}, \citenamefont {Leproux},\ and\ \citenamefont
  {Couderc}}]{Baronio12}%
  \BibitemOpen
  \bibfield  {author} {\bibinfo {author} {\bibfnamefont {F.}~\bibnamefont
  {Baronio}}, \bibinfo {author} {\bibfnamefont {M.}~\bibnamefont {Conforti}},
  \bibinfo {author} {\bibfnamefont {C.}~\bibnamefont {{De Angelis}}}, \bibinfo
  {author} {\bibfnamefont {D.}~\bibnamefont {Modotto}}, \bibinfo {author}
  {\bibfnamefont {S.}~\bibnamefont {Wabnitz}}, \bibinfo {author} {\bibfnamefont
  {M.}~\bibnamefont {Andreana}}, \bibinfo {author} {\bibfnamefont
  {A.}~\bibnamefont {Tonello}}, \bibinfo {author} {\bibfnamefont
  {P.}~\bibnamefont {Leproux}},\ and\ \bibinfo {author} {\bibfnamefont
  {V.}~\bibnamefont {Couderc}},\ }\bibfield  {title} {\bibinfo {title} {Second
  and third order susceptibilities mixing for supercontinuum generation and
  shaping},\ }\href
  {https://doi.org/https://doi.org/10.1016/j.yofte.2012.07.001} {\bibfield
  {journal} {\bibinfo  {journal} {Optical Fiber Technology}\ }\textbf {\bibinfo
  {volume} {18}},\ \bibinfo {pages} {283} (\bibinfo {year} {2012})},\ \bibinfo
  {note} {fiber Supercontinuum sources and their applications}\BibitemShut
  {NoStop}%
\bibitem [{\citenamefont {Wabnitz}\ and\ \citenamefont
  {Kozlov}(2010)}]{Wabnitz10}%
  \BibitemOpen
  \bibfield  {author} {\bibinfo {author} {\bibfnamefont {S.}~\bibnamefont
  {Wabnitz}}\ and\ \bibinfo {author} {\bibfnamefont {V.~V.}\ \bibnamefont
  {Kozlov}},\ }\bibfield  {title} {\bibinfo {title} {Harmonic and
  supercontinuum generation in quadratic and cubic nonlinear optical media},\
  }\href {https://doi.org/10.1364/JOSAB.27.001707} {\bibfield  {journal}
  {\bibinfo  {journal} {J. Opt. Soc. Am. B}\ }\textbf {\bibinfo {volume}
  {27}},\ \bibinfo {pages} {1707} (\bibinfo {year} {2010})}\BibitemShut
  {NoStop}%
\bibitem [{\citenamefont {Phillips}\ \emph {et~al.}(2011)\citenamefont
  {Phillips}, \citenamefont {Langrock}, \citenamefont {Pelc}, \citenamefont
  {Fejer}, \citenamefont {Hartl},\ and\ \citenamefont {Fermann}}]{Phillips11}%
  \BibitemOpen
  \bibfield  {author} {\bibinfo {author} {\bibfnamefont {C.~R.}\ \bibnamefont
  {Phillips}}, \bibinfo {author} {\bibfnamefont {C.}~\bibnamefont {Langrock}},
  \bibinfo {author} {\bibfnamefont {J.~S.}\ \bibnamefont {Pelc}}, \bibinfo
  {author} {\bibfnamefont {M.~M.}\ \bibnamefont {Fejer}}, \bibinfo {author}
  {\bibfnamefont {I.}~\bibnamefont {Hartl}},\ and\ \bibinfo {author}
  {\bibfnamefont {M.~E.}\ \bibnamefont {Fermann}},\ }\bibfield  {title}
  {\bibinfo {title} {Supercontinuum generation in quasi-phasematched
  waveguides},\ }\href {https://doi.org/10.1364/OE.19.018754} {\bibfield
  {journal} {\bibinfo  {journal} {Opt. Express}\ }\textbf {\bibinfo {volume}
  {19}},\ \bibinfo {pages} {18754} (\bibinfo {year} {2011})}\BibitemShut
  {NoStop}%
\bibitem [{\citenamefont {Conforti}\ \emph {et~al.}(2010)\citenamefont
  {Conforti}, \citenamefont {Baronio},\ and\ \citenamefont
  {De~Angelis}}]{Conforti10}%
  \BibitemOpen
  \bibfield  {author} {\bibinfo {author} {\bibfnamefont {M.}~\bibnamefont
  {Conforti}}, \bibinfo {author} {\bibfnamefont {F.}~\bibnamefont {Baronio}},\
  and\ \bibinfo {author} {\bibfnamefont {C.}~\bibnamefont {De~Angelis}},\
  }\bibfield  {title} {\bibinfo {title} {Nonlinear envelope equation for
  broadband optical pulses in quadratic media},\ }\href
  {https://doi.org/10.1103/PhysRevA.81.053841} {\bibfield  {journal} {\bibinfo
  {journal} {Phys. Rev. A}\ }\textbf {\bibinfo {volume} {81}},\ \bibinfo
  {pages} {053841} (\bibinfo {year} {2010})}\BibitemShut {NoStop}%
\bibitem [{\citenamefont {Kolesik}\ and\ \citenamefont
  {Moloney}(2004)}]{Kolesik04}%
  \BibitemOpen
  \bibfield  {author} {\bibinfo {author} {\bibfnamefont {M.}~\bibnamefont
  {Kolesik}}\ and\ \bibinfo {author} {\bibfnamefont {J.~V.}\ \bibnamefont
  {Moloney}},\ }\bibfield  {title} {\bibinfo {title} {Nonlinear optical pulse
  propagation simulation: From maxwell's to unidirectional equations},\ }\href
  {https://doi.org/10.1103/PhysRevE.70.036604} {\bibfield  {journal} {\bibinfo
  {journal} {Phys. Rev. E}\ }\textbf {\bibinfo {volume} {70}},\ \bibinfo
  {pages} {036604} (\bibinfo {year} {2004})}\BibitemShut {NoStop}%
\bibitem [{\citenamefont {Conforti}\ \emph {et~al.}(2013)\citenamefont
  {Conforti}, \citenamefont {Marini}, \citenamefont {Tran}, \citenamefont
  {Faccio},\ and\ \citenamefont {Biancalana}}]{Conforti13}%
  \BibitemOpen
  \bibfield  {author} {\bibinfo {author} {\bibfnamefont {M.}~\bibnamefont
  {Conforti}}, \bibinfo {author} {\bibfnamefont {A.}~\bibnamefont {Marini}},
  \bibinfo {author} {\bibfnamefont {T.~X.}\ \bibnamefont {Tran}}, \bibinfo
  {author} {\bibfnamefont {D.}~\bibnamefont {Faccio}},\ and\ \bibinfo {author}
  {\bibfnamefont {F.}~\bibnamefont {Biancalana}},\ }\bibfield  {title}
  {\bibinfo {title} {Interaction between optical fields and their conjugates in
  nonlinear media},\ }\href {https://doi.org/10.1364/OE.21.031239} {\bibfield
  {journal} {\bibinfo  {journal} {Opt. Express}\ }\textbf {\bibinfo {volume}
  {21}},\ \bibinfo {pages} {31239} (\bibinfo {year} {2013})}\BibitemShut
  {NoStop}%
\bibitem [{\citenamefont {Agrawal}(2007)}]{Agrawal07}%
  \BibitemOpen
  \bibfield  {author} {\bibinfo {author} {\bibfnamefont {G.~P.}\ \bibnamefont
  {Agrawal}},\ }\href@noop {} {\emph {\bibinfo {title} {Nonlinear Fiber
  Optics}}},\ \bibinfo {edition} {4th}\ ed.,\ San Diego, California\ (\bibinfo
  {publisher} {Academic Press},\ \bibinfo {year} {2007})\BibitemShut {NoStop}%
\bibitem [{\citenamefont {Guo}\ \emph {et~al.}(2018)\citenamefont {Guo},
  \citenamefont {Herkommer}, \citenamefont {Billat}, \citenamefont {Grassani},
  \citenamefont {Zhang}, \citenamefont {Pfeiffer}, \citenamefont {Weng},
  \citenamefont {Br{\`e}s},\ and\ \citenamefont {Kippenberg}}]{Guo18}%
  \BibitemOpen
  \bibfield  {author} {\bibinfo {author} {\bibfnamefont {H.}~\bibnamefont
  {Guo}}, \bibinfo {author} {\bibfnamefont {C.}~\bibnamefont {Herkommer}},
  \bibinfo {author} {\bibfnamefont {A.}~\bibnamefont {Billat}}, \bibinfo
  {author} {\bibfnamefont {D.}~\bibnamefont {Grassani}}, \bibinfo {author}
  {\bibfnamefont {C.}~\bibnamefont {Zhang}}, \bibinfo {author} {\bibfnamefont
  {M.~H.~P.}\ \bibnamefont {Pfeiffer}}, \bibinfo {author} {\bibfnamefont
  {W.}~\bibnamefont {Weng}}, \bibinfo {author} {\bibfnamefont {C.-S.}\
  \bibnamefont {Br{\`e}s}},\ and\ \bibinfo {author} {\bibfnamefont {T.~J.}\
  \bibnamefont {Kippenberg}},\ }\bibfield  {title} {\bibinfo {title}
  {Mid-infrared frequency comb via coherent dispersive wave generation in
  silicon nitride nanophotonic waveguides},\ }\href
  {https://doi.org/10.1038/s41566-018-0144-1} {\bibfield  {journal} {\bibinfo
  {journal} {Nature Photonics}\ }\textbf {\bibinfo {volume} {12}},\ \bibinfo
  {pages} {330} (\bibinfo {year} {2018})}\BibitemShut {NoStop}%
\bibitem [{\citenamefont {Zhao}\ \emph {et~al.}(2020)\citenamefont {Zhao},
  \citenamefont {Ji}, \citenamefont {Kim}, \citenamefont {Donvalkar},
  \citenamefont {Jang}, \citenamefont {Joshi}, \citenamefont {Yu},
  \citenamefont {Joshi}, \citenamefont {Domeneguetti}, \citenamefont {Barbosa},
  \citenamefont {Nussenzveig}, \citenamefont {Okawachi}, \citenamefont
  {Lipson},\ and\ \citenamefont {Gaeta}}]{Zhao20}%
  \BibitemOpen
  \bibfield  {author} {\bibinfo {author} {\bibfnamefont {Y.}~\bibnamefont
  {Zhao}}, \bibinfo {author} {\bibfnamefont {X.}~\bibnamefont {Ji}}, \bibinfo
  {author} {\bibfnamefont {B.~Y.}\ \bibnamefont {Kim}}, \bibinfo {author}
  {\bibfnamefont {P.~S.}\ \bibnamefont {Donvalkar}}, \bibinfo {author}
  {\bibfnamefont {J.~K.}\ \bibnamefont {Jang}}, \bibinfo {author}
  {\bibfnamefont {C.}~\bibnamefont {Joshi}}, \bibinfo {author} {\bibfnamefont
  {M.}~\bibnamefont {Yu}}, \bibinfo {author} {\bibfnamefont {C.}~\bibnamefont
  {Joshi}}, \bibinfo {author} {\bibfnamefont {R.~R.}\ \bibnamefont
  {Domeneguetti}}, \bibinfo {author} {\bibfnamefont {F.~A.~S.}\ \bibnamefont
  {Barbosa}}, \bibinfo {author} {\bibfnamefont {P.}~\bibnamefont
  {Nussenzveig}}, \bibinfo {author} {\bibfnamefont {Y.}~\bibnamefont
  {Okawachi}}, \bibinfo {author} {\bibfnamefont {M.}~\bibnamefont {Lipson}},\
  and\ \bibinfo {author} {\bibfnamefont {A.~L.}\ \bibnamefont {Gaeta}},\
  }\bibfield  {title} {\bibinfo {title} {Visible nonlinear photonics via
  high-order-mode dispersion engineering},\ }\href
  {https://doi.org/10.1364/OPTICA.7.000135} {\bibfield  {journal} {\bibinfo
  {journal} {Optica}\ }\textbf {\bibinfo {volume} {7}},\ \bibinfo {pages} {135}
  (\bibinfo {year} {2020})}\BibitemShut {NoStop}%
\end{thebibliography}
%


\end{document}